\documentstyle[twoside,fleqn,espcrc2]{article}
\input prepictex
\input pictex
\input postpictex

\newdimen\xposition
\newdimen\yposition
\newdimen\dyposition
\newdimen\crossbarlength

\def\putcircbar at #1 #2 with fuzz #3 {%
  \xposition=\Xdistance{#1}
  \yposition=\Ydistance{#2}
  \dyposition=\Ydistance{#3}

\setdimensionmode
\put {$\circ$} at {\xposition} {\yposition}

\dimen0 = \yposition
  \advance \dimen0 by -\dyposition
\dimen2 = \yposition
  \advance \dimen2  by \dyposition
\putrule from {\xposition} {\dimen0}
  to {\xposition} {\dimen2}

\dimen4 = \xposition
  \advance \dimen4 by -.5\crossbarlength
\dimen6 = \xposition
  \advance \dimen6 by  .5\crossbarlength
\putrule from {\dimen4} {\dimen0} to {\dimen6} {\dimen0}
\putrule from {\dimen4} {\dimen2} to {\dimen6} {\dimen2}
\setcoordinatemode}

\newdimen\xposition
\newdimen\yposition
\newdimen\dyposition
\newdimen\crossbarlength

\def\putdiamondbar at #1 #2 with fuzz #3 {%
  \xposition=\Xdistance{#1}
  \yposition=\Ydistance{#2}
  \dyposition=\Ydistance{#3}

\setdimensionmode
\put {$\diamond$} at {\xposition} {\yposition}

\dimen0 = \yposition
  \advance \dimen0 by -\dyposition
\dimen2 = \yposition
  \advance \dimen2  by \dyposition
\putrule from {\xposition} {\dimen0}
  to {\xposition} {\dimen2}

\dimen4 = \xposition
  \advance \dimen4 by -.5\crossbarlength
\dimen6 = \xposition
  \advance \dimen6 by  .5\crossbarlength
\putrule from {\dimen4} {\dimen0} to {\dimen6} {\dimen0}
\putrule from {\dimen4} {\dimen2} to {\dimen6} {\dimen2}
\setcoordinatemode}

\newdimen\xposition
\newdimen\yposition
\newdimen\dyposition
\newdimen\crossbarlength

\def\putbigtriangledownbar at #1 #2 with fuzz #3 {%
  \xposition=\Xdistance{#1}
  \yposition=\Ydistance{#2}
  \dyposition=\Ydistance{#3}

\setdimensionmode
\put {\tiny $\bigtriangledown$} at {\xposition} {\yposition}

\dimen0 = \yposition
  \advance \dimen0 by -\dyposition
\dimen2 = \yposition
  \advance \dimen2  by \dyposition
\putrule from {\xposition} {\dimen0}
  to {\xposition} {\dimen2}

\dimen4 = \xposition
  \advance \dimen4 by -.5\crossbarlength
\dimen6 = \xposition
  \advance \dimen6 by  .5\crossbarlength
\putrule from {\dimen4} {\dimen0} to {\dimen6} {\dimen0}
\putrule from {\dimen4} {\dimen2} to {\dimen6} {\dimen2}
\setcoordinatemode}

\newdimen\xposition
\newdimen\yposition
\newdimen\dyposition
\newdimen\crossbarlength

\def\puttrianglebar at #1 #2 with fuzz #3 {%
  \xposition=\Xdistance{#1}
  \yposition=\Ydistance{#2}
  \dyposition=\Ydistance{#3}

\setdimensionmode
\put {\tiny $\triangle$} at {\xposition} {\yposition}

\dimen0 = \yposition
  \advance \dimen0 by -\dyposition
\dimen2 = \yposition
  \advance \dimen2  by \dyposition
\putrule from {\xposition} {\dimen0}
  to {\xposition} {\dimen2}

\dimen4 = \xposition
  \advance \dimen4 by -.5\crossbarlength
\dimen6 = \xposition
  \advance \dimen6 by  .5\crossbarlength
\putrule from {\dimen4} {\dimen0} to {\dimen6} {\dimen0}
\putrule from {\dimen4} {\dimen2} to {\dimen6} {\dimen2}
\setcoordinatemode}

\newdimen\xposition
\newdimen\yposition
\newdimen\dyposition
\newdimen\crossbarlength

\def\puttrianglerightbar at #1 #2 with fuzz #3 {%
  \xposition=\Xdistance{#1}
  \yposition=\Ydistance{#2}
  \dyposition=\Ydistance{#3}

\setdimensionmode
\put {$\triangleright$} at {\xposition} {\yposition}

\dimen0 = \yposition
  \advance \dimen0 by -\dyposition
\dimen2 = \yposition
  \advance \dimen2  by \dyposition
\putrule from {\xposition} {\dimen0}
  to {\xposition} {\dimen2}

\dimen4 = \xposition
  \advance \dimen4 by -.5\crossbarlength
\dimen6 = \xposition
  \advance \dimen6 by  .5\crossbarlength
\putrule from {\dimen4} {\dimen0} to {\dimen6} {\dimen0}
\putrule from {\dimen4} {\dimen2} to {\dimen6} {\dimen2}
\setcoordinatemode}

\newdimen\xposition
\newdimen\yposition
\newdimen\dyposition

\def\puttriangleleftbar at #1 #2 with fuzz #3 {%
  \xposition=\Xdistance{#1}
  \yposition=\Ydistance{#2}
  \dyposition=\Ydistance{#3}

\setdimensionmode
\put {$\triangleleft$} at {\xposition} {\yposition}

\dimen0 = \yposition
  \advance \dimen0 by -\dyposition
\dimen2 = \yposition
  \advance \dimen2  by \dyposition
\putrule from {\xposition} {\dimen0}
  to {\xposition} {\dimen2}

\dimen4 = \xposition
  \advance \dimen4 by -.5\crossbarlength
\dimen6 = \xposition
  \advance \dimen6 by  .5\crossbarlength
\putrule from {\dimen4} {\dimen0} to {\dimen6} {\dimen0}
\putrule from {\dimen4} {\dimen2} to {\dimen6} {\dimen2}
\setcoordinatemode}

\newcommand{\AmS}{{\protect\the\textfont2
  A\kern-.1667em\lower.5ex\hbox{M}\kern-.125emS}}

\title{Noncompact Gauge-Invariant Simulations of $U(1)$, $SU(2)$, and $SU(3)$}

\author{K.~Cahill\thanks{Research supported by the
U.~S. Department of Energy; 
e-mail: kevin @ cahill.phys.unm.edu.}
and G.~Herling\thanks{Member of the Center for Advanced Studies;
e-mail: herling @ bootes.unm.edu.}\\
{ \hskip 1cm } \\
Department of Physics and Astronomy, University of New Mexico\\
Albuquerque, New Mexico 87131-1156, U.~S.~A.}

\begin{document}

\begin{abstract}
We have applied a new gauge-invariant, noncompact,
Monte Carlo method to simulate
the $U(1)$, $SU(2)$, and $SU(3)$ gauge theories
on $8^4$ and $12^4$ lattices.
The Creutz ratios of the Wilson loops
agree with the exact results for $U(1)$
for $\beta \ge 0.5$
apart from a renormalization of the charge.
The $SU(2)$ and $SU(3)$ Creutz ratios
robustly display quark confinement at $\beta = 0.5$
and $\beta = 2$, respectively.
At much weaker coupling, the $SU(2)$ and $SU(3)$ Creutz ratios
agree with perturbation theory after a renormalization
of the coupling constant.
For $SU(3)$ the scaling window
is near $ \beta = 2 $,
and the relation between the string tension $\sigma $
and our lattice QCD parameter $ \Lambda_L $
is $\sqrt{\sigma} \approx 5 \Lambda_L$.
\end{abstract}

\maketitle

\section{INTRODUCTION}

In compact lattice gauge theory,
gauge fields are represented by group elements
rather than by fields,
and the action is a {\it periodic\/} function
of a gauge-invariant lattice field strength.
The periodicity of the action entails spurious vacua.
The principal advantage of noncompact actions,
in which gauge fields are represented by fields,
is that they avoid multiple vacua.
\par
The first gauge-invariant noncompact simulations
were carried out by Palumbo, Polikarpov, and Veselov~\cite{Palumbo92}.
They saw a confinement signal.
Their action contains five terms,
constructed from two invariants,
and involves (noncompact) auxiliary fields
and an adjustable parameter.
\par
The present paper describes a test
of a new way~\cite{CahGary}
of performing gauge-invariant
noncompact simulations.
Our action, which is
similar to one term
of Palumbo's action,
is exactly invariant under compact
gauge transformations,
is a natural discretization
of the classical Yang-Mills action,
and reduces to Wilson's action
when the gauge fields are compactified.
In this method
there are fewer auxiliary fields
than in Palumbo's method,
and they are compact group elements representing
gauge transformations.
\par
We have used this method
to simulate $U(1)$, $SU(2)$, and $SU(3)$ gauge theories
on $8^4$ and $12^4$ lattices.
The Creutz ratios of Wilson loops
agree with the exact results for $U(1)$
for $\beta \ge 0.5$
apart from a renormalization of the charge.
The $SU(2)$ and $SU(3)$ Creutz ratios
clearly show quark confinement at $\beta = 0.5$
and $\beta = 2$, respectively.
At much weaker coupling, the $SU(2)$ and $SU(3)$ Creutz ratios
agree with perturbation theory with a renormalized
coupling constant.
For $SU(3)$ there is a scaling window near $\beta=2$,
and the string tension $\sigma $
is related to the lattice QCD parameter $ \Lambda_L $
by $\sqrt{\sigma} \approx 5 \Lambda_L$.
If $\sqrt{\sigma} \approx 420$ MeV,
then our $ \Lambda_L $ is about 84 MeV,
and at $\beta = 2$ our lattice spacing $a$
is about 0.4 fm.


\section{THE METHOD}

For massless fermions, the continuum action density is
$\bar \psi i \gamma_\mu \partial_\mu \psi.$
A suitable discretization of this quantity is
$i \bar \psi (n) \gamma_\mu
[ \psi(n + e_\mu) - \psi(n)]/a$
in which $n$ is a four-vector
of integers representing an arbitrary vertex
of the lattice, $e_\mu$ is a unit vector
in the $\mu$th direction, and
$a$ is the lattice spacing.
The product of Fermi fields at the same
point is gauge invariant as it stands.
The other product of Fermi fields
becomes gauge invariant
if we insert a matrix $A_\mu(n)$ of gauge fields
\begin{equation}
\bar \psi (n) \gamma_\mu
\left[ 1 + i g a A_\mu(n) \right] \psi(n + e_\mu)
\end{equation}
that transforms appropriately.
Under a gauge transformation
represented by the group elements
$U(n)$ and $U(n + e_\mu)$,
the required response is
\begin{equation}
1 + i a g A'_\mu(n) =
U(n) [ 1 + i a g A_\mu(n) ] U^{-1}(n + e_\mu).
\label{A'}
\end{equation}
Under this gauge transformation,
the lattice field strength
\begin{eqnarray}
F_{\mu\nu}(n) & = &
{ 1 \over a } \left[ A_\mu(n+e_\nu) - A_\mu(n) \right]
\nonumber \\
& & \mbox{}
- { 1 \over a } \left[ A_\nu(n+e_\mu) - A_\nu(n) \right]
\nonumber \\
& & \mbox{} + i g  \left[ A_\nu(n) A_\mu(n+e_\nu) \right.
\nonumber \\
& & \mbox{}
\quad \left. - A_\mu(n) A_\nu(n+e_\mu) \right],
\label{F}
\end{eqnarray}
which reduces to the continuum Yang-Mills
field strength in the limit $a \to 0$,
transforms as
\begin{equation}
F'_{\mu\nu}(n) = U(n) F_{\mu\nu}(n) U^{-1}(n + e_\mu + e_\nu).
\label{F'}
\end{equation}
The field strength $F_{\mu\nu}(n)$
is antisymmetric
in the indices $\mu$ and $\nu$, but it is not
hermitian.
To make a positive plaquette action density,
we use the Hilbert-Schmidt norm of $F_{\mu\nu}(n)$
\begin{equation}
S = {1 \over 4 k} {\rm Tr} [F^\dagger_{\mu\nu}(n) F_{\mu\nu}(n)],
\label{S}
\end{equation}
in which the generators $T_a$
of the gauge group are normalized as
${\rm Tr} ( T_a T_b ) = k \delta_{ab}$\null.
Because $F_{\mu\nu}(n)$ transforms
covariantly (\ref{F'}), this action density
is exactly invariant under the
noncompact gauge transformation (\ref{A'}).
\par
In general
the gauge transformation (\ref{A'})
with group element $U(n) = \exp(-i a g \omega^a T_a )$
maps the matrix of gauge fields
$A_\mu(n) = T_a A^a_\mu(n)$
outside the Lie algebra,
apart from terms of lowest (zeroth) order in the
lattice spacing $a$\null.
We use this larger space of matrices.
We use the action (\ref{S})
in which the field strength (\ref{F})
is defined in terms of gauge-field
matrices $ A_\mu(n) $ that are
the images under arbitrary gauge
transformations
\begin{equation}
A_\mu(n) = V A^0_\mu(n) W^{-1}
+ {i \over a g} V \left( V^{-1} - W^{-1} \right)
\label{newA'}
\end{equation}
of matrices $ A^0_\mu(n) $ of gauge fields
defined in the usual way,
$ A^0_\mu(n) \equiv T_a A^{a,0}_\mu(n) $.
The group elements $V$ and $W$
associated with the gauge field $A_\mu(n)$
are unrelated to those
associated with the neighboring
gauge fields $ A_\mu(n+e_\nu) $,
$ A_\nu(n) $, and $ A_\nu(n+e_\mu) $.
\par
The quantity
$ 1 + iga A_\mu(n) $
is not an element $ L_\mu(n) $
of the gauge group.
But if one compactified
the fields by requiring $ 1 + iga A_\mu(n) $
to be an element of the gauge group,
then the matrix $ A_\mu(n) $
of gauge fields would be
related to the link $ L_\mu(n) $ by
$A_\mu(n) =  ( L_\mu(n) - 1 )/( iga )$,
and the action (\ref{S})
defined in terms of the field strength
(\ref{F}) would be, {\it mirabile dictu\/},
Wilson's action:
$$
S =
{k - \Re \, {\rm Tr} L_\mu(n) L_\nu(n+e_\mu)
L^\dagger_\mu(n+e_\nu) L^\dagger_\nu(n)
\over 2 a^4 g^2k }.
$$

\section{RESULTS}

We have tested this method by applying it
to the $U(1)$, $SU(2)$, and $SU(3)$ gauge
theories on $8^4$ and $12^4$ lattices.
In most of our initial configurations,
the unitary matrices $V$ and $W$
and the hermitian gauge fields $A^0_\mu$
were randomized.
For thermalization
we allowed 50,000 sweeps for $U(1)$,
10,000 for $SU(2)$, and 100,000
for $SU(3)$.
Our Wilson loops are ensemble averages
of ordered products of the binomials
$1 + i a g A_\mu(n)$
rather than of the exponentials
$\exp[i a g A_\mu(n)]$ around the loop.
\par
For $U(1)$ and for $\beta = 1/g^2 \ge 0.5$,
our measured Creutz ratios~\cite{Creu80b}
of Wilson loops
agree with the exact ones
apart from finite-size effects
and a renormalization of the charge.
For instance at $\beta = 0.5$
on the $12^4$ lattice, we found
$\chi(2,2) =  0.147(1)$,
$\chi(2,3) =  0.103(1)$,
$\chi(2,4) =  0.090(1)$,
$\chi(3,3) =  0.049(1)$,
$\chi(3,4) =  0.034(1)$, and
$\chi(4,4) =  0.020(2)$.
The first three of these $\chi$'s are equal to the
exact Creutz ratios for a renormalized value of
$\beta_r = 0.45 $; the last three are
smaller than the exact ratios for $\beta_r = 0.45 $
due to finite-size effects by 3\%, 8\%, and 16\%, respectively.
\par
But at stronger coupling,
the extra terms
$i g  \left[ A_\nu(n) A_\mu(n+e_\nu)
- A_\mu(n) A_\nu(n+e_\mu) \right]$
in the lattice field
strength $F_{\mu\nu}(n)$
eventually do produce a
confinement signal.
For example, at $\beta = 0.375 $,
our measured Creutz ratios
on the $12^4$ lattice are:
$\chi(2,2) =  0.906(5)$,
$\chi(2,3) =  0.909(21)$,
$\chi(2,4) =  0.85(10)$,
$\chi(3,3) =  0.62(24)$, and
$\chi(3,4) =  0.6(16)$.
\par
For $SU(2)$
on the $8^4$ lattice at $\beta = 4/g^2 = 0.5$,
we found
$\chi(2,2) =  0.835(3)$,
$\chi(2,3) =  0.852(12)$,
$\chi(2,4) =  0.865(60)$, and
$\chi(3,3) =  0.94(23)$
which within the limited statistics
clearly exhibit confinement.
At $\beta = 1$, our six Creutz ratios
track those of tree-level perturbation
theory for a renormalized value of $\beta_r = 1.75$.
\par
For $SU(3)$ at $\beta = 6/g^2 = 2$
on the $12^4$ lattice,
we found in ten independent runs
$\chi(2,2) =  0.838(1)$,
$\chi(2,3) =  0.826(3)$,
$\chi(2,4) =  0.828(13)$,
$\chi(3,3) =  0.793(42)$,
$\chi(3,4) =  0.47(25)$, and
$\chi(4,4) =  1.2(86)$.
Within the statistics,
these results robustly exhibit confinement.
At much weaker coupling, our ratios
agree with perturbation theory
apart from finite-size effects and
after a renormalization of the coupling constant.

\section{SCALING}
\par
We used an $8^4$ lattice to
study the scaling of the lattice
spacing $a$ with the coupling constant $g$
for $SU(3)$.  The two-loop result
for the dependence of the string tension
$\sigma a^2$ upon the inverse coupling
$\beta = 6/g^2$ is
\begin{equation}
\sigma a^2 \approx
{ \sigma \over \Lambda_L^2 }
\exp \left[ - { 8 \pi^2 \beta \over 33 }
+ {102 \over 121} \log\left( { 8 \pi^2 \beta \over 33 } \right)
\right].
\label{scaling}
\end{equation}
If we set $ \sqrt{\sigma} \approx (5.0 \pm 0.4) \Lambda_L$,
then our $\chi(i,j)$'s
fit this formula
for $ 1.9 < \beta < 2.1 $
as shown in the figure.
A string tension
$\sqrt{\sigma} \approx 420$ MeV
implies that
$ \Lambda_L \approx 84$ MeV,
which is about 11 times
closer to the the continuum
$ \Lambda_{\overline{\rm MS}}^{(0)} $
than is the parameter $ \Lambda_{LW} \approx 7.9$ MeV
of Wilson's method.
At $\beta = 2$, our lattice spacing $a$
is about 0.4 fm.  

\begin{figure} [htb]
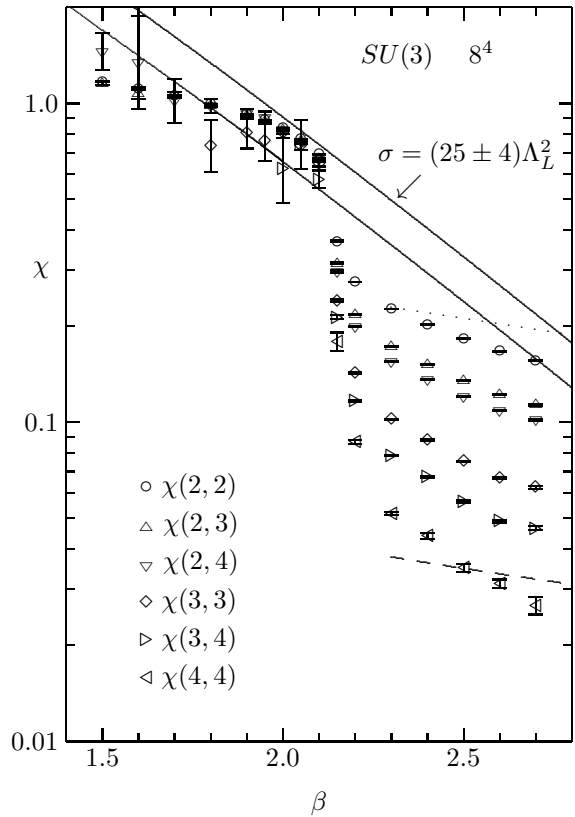

\beginpicture
\crossbarlength=5pt

\ticksin
\inboundscheckon
\setcoordinatesystem units <1.89in,1.671in>   
\setplotarea x from 1.4 to 2.8, y from -2 to 0.30103
\axis bottom label {$\beta$} ticks
  numbered at 1.5 2.0 2.5 /
  unlabeled short from 1.6 to 1.9 by 0.1
  from 2.1 to 2.4 by 0.1 from 2.6 to 2.7 by 0.1 /
\axis left 
  ticks logged
  numbered at 0.01 0.1 1.0 /
  unlabeled short at 2.0 /
  withvalues {$\chi$ } /
  at 0.3 /
  unlabeled short from 0.2 to 0.9 by 0.1
  from 0.02 to 0.09 by 0.01 /
\axis top label {}
  ticks
  unlabeled at 1.5 2.0 2.5 /
  unlabeled short from 1.6 to 1.9 by 0.1
  from 2.1 to 2.4 by 0.1 from 2.6 to 2.7 by 0.1 /
\axis right ticks logged
  unlabeled at 0.01 0.1 1.0 /
  unlabeled short at 2.0 /
  unlabeled short from 0.2 to 0.9 by 0.1
  from 0.02 to 0.09 by 0.01 /

\put {$SU(3)$ \quad $8^4$} at 2.4 0.15

\setquadratic
\plot 1.4 0.310035 1.8 -0.013601 2.0 -0.182850 2.4 -0.531745 2.8 -0.890953 /
\plot 1.58 0.307455 1.8 0.126577 2.0 -0.042671 2.4 -0.391566 2.8 -0.750774 /
\put {$\swarrow$} at 2.35 -0.26
\put {$\sigma = (25 \pm 4)\Lambda_L^2$} at 2.51 -0.16

\inboundscheckon

\putcircbar at  1.500  0.067265 with fuzz  0.000993
\puttrianglebar at  1.500  0.063200 with fuzz  0.005738
\putbigtriangledownbar at  1.500  0.162397 with fuzz  0.057040

\putcircbar at  1.600  0.046057 with fuzz  0.004179
\puttrianglebar at  1.600  0.030172 with fuzz  0.015589
\putbigtriangledownbar at  1.600  0.127390 with fuzz  0.146388

\putcircbar at  1.700  0.024673 with fuzz  0.002118
\puttrianglebar at  1.700  0.026838 with fuzz  0.008122
\putbigtriangledownbar at  1.700  0.007245 with fuzz  0.069343

\putcircbar at  1.800 -0.001990 with fuzz  0.000861
\puttrianglebar at  1.800 -0.006465 with fuzz  0.003713
\putdiamondbar at  1.800 -0.133903 with fuzz  0.082046
\putbigtriangledownbar at  1.800 -0.008941 with fuzz  0.022102

\putcircbar at  1.900 -0.034160 with fuzz  0.000751
\puttrianglebar at  1.900 -0.039606 with fuzz  0.002866
\putdiamondbar at  1.900 -0.094664 with fuzz  0.047367
\putbigtriangledownbar at  1.900 -0.032220 with fuzz  0.013574

\putcircbar at  1.950 -0.054024 with fuzz  0.001125
\puttrianglebar at  1.950 -0.058458 with fuzz  0.003958
\putdiamondbar at  1.950 -0.118123 with fuzz  0.063084
\putbigtriangledownbar at  1.950 -0.044281 with fuzz  0.019162

\putcircbar at  2.000 -0.078531 with fuzz  0.000341
\puttrianglebar at  2.000 -0.083955 with fuzz  0.000999
\putdiamondbar at  2.000 -0.094706 with fuzz  0.013256
\putbigtriangledownbar at  2.000 -0.083701 with fuzz  0.004086
\puttrianglerightbar at  2.000 -0.204388 with fuzz  0.107113

\putcircbar at  2.050 -0.111724 with fuzz  0.000605
\puttrianglebar at  2.050 -0.118746 with fuzz  0.001244
\putdiamondbar at  2.050 -0.133833 with fuzz  0.013837
\putbigtriangledownbar at  2.050 -0.121984 with fuzz  0.004430
\puttrianglerightbar at  2.050 -0.128772 with fuzz  0.076798

\putcircbar at  2.100 -0.160306 with fuzz  0.000423
\puttrianglebar at  2.100 -0.173155 with fuzz  0.000684
\putdiamondbar at  2.100 -0.191054 with fuzz  0.007317
\putbigtriangledownbar at  2.100 -0.177235 with fuzz  0.001802
\puttrianglerightbar at  2.100 -0.238247 with fuzz  0.026799

\putcircbar at  2.150 -0.433324 with fuzz  0.002403
\puttrianglebar at  2.150 -0.501220 with fuzz  0.003081
\putdiamondbar at  2.150 -0.619365 with fuzz  0.004751
\putbigtriangledownbar at  2.150 -0.527851 with fuzz  0.003441
\puttrianglerightbar at  2.150 -0.670727 with fuzz  0.006297
\puttriangleleftbar at  2.150 -0.747661 with fuzz  0.028757

\putcircbar at  2.200 -0.560792 with fuzz  0.000508
\puttrianglebar at  2.200 -0.661330 with fuzz  0.000763
\putdiamondbar at  2.200 -0.846271 with fuzz  0.001665
\putbigtriangledownbar at  2.200 -0.698813 with fuzz  0.000947
\puttrianglerightbar at  2.200 -0.933111 with fuzz  0.002544
\puttriangleleftbar at  2.200 -1.061830 with fuzz  0.006592

\putcircbar at  2.300 -0.644322 with fuzz  0.000281
\puttrianglebar at  2.300 -0.763192 with fuzz  0.000415
\putdiamondbar at  2.300 -0.989687 with fuzz  0.000966
\putbigtriangledownbar at  2.300 -0.808837 with fuzz  0.000521
\puttrianglerightbar at  2.300 -1.103702 with fuzz  0.001463
\puttriangleleftbar at  2.300 -1.286620 with fuzz  0.004451

\putcircbar at  2.400 -0.694850 with fuzz  0.000598
\puttrianglebar at  2.400 -0.819663 with fuzz  0.000895
\putdiamondbar at  2.400 -1.055979 with fuzz  0.002571
\putbigtriangledownbar at  2.400 -0.865585 with fuzz  0.001222
\puttrianglerightbar at  2.400 -1.171882 with fuzz  0.003933
\puttriangleleftbar at  2.400 -1.356133 with fuzz  0.010208

\putcircbar at  2.500 -0.738753 with fuzz  0.000621
\puttrianglebar at  2.500 -0.869129 with fuzz  0.000927
\putdiamondbar at  2.500 -1.122638 with fuzz  0.002250
\putbigtriangledownbar at  2.500 -0.919629 with fuzz  0.001221
\puttrianglerightbar at  2.500 -1.249248 with fuzz  0.003605
\puttriangleleftbar at  2.500 -1.456392 with fuzz  0.012055

\putcircbar at  2.600 -0.777804 with fuzz  0.000755
\puttrianglebar at  2.600 -0.914167 with fuzz  0.001115
\putdiamondbar at  2.600 -1.174866 with fuzz  0.003239
\putbigtriangledownbar at  2.600 -0.963280 with fuzz  0.001459
\puttrianglerightbar at  2.600 -1.309962 with fuzz  0.004209
\puttriangleleftbar at  2.600 -1.505928 with fuzz  0.013414

\putcircbar at  2.700 -0.807103 with fuzz  0.001772
\puttrianglebar at  2.700 -0.944852 with fuzz  0.002659
\putdiamondbar at  2.700 -1.204428 with fuzz  0.003996
\putbigtriangledownbar at  2.700 -0.992567 with fuzz  0.003285
\puttrianglerightbar at  2.700 -1.332004 with fuzz  0.006333
\puttriangleleftbar at  2.700 -1.574826 with fuzz  0.027404

\put {$\circ$  $\chi(2,2)$ } at 1.75 -1.2
\put {{\tiny $\triangle$}  $\chi(2,3)$ } at 1.75 -1.32
\put {{\tiny $\bigtriangledown$}  $\chi(2,4)$ } at 1.75 -1.44
\put {$\diamond$  $\chi(3,3)$ } at 1.75 -1.56
\put {$\triangleright$  $\chi(3,4)$ } at 1.75 -1.68
\put {$\triangleleft$  $\chi(4,4)$ } at 1.75 -1.80

\setdots
\setquadratic 
\plot
2.3 -0.638569211022062
2.4 -0.657052616716075
2.5 -0.674781383676507
2.6 -0.691814722975287
2.7 -0.708205139163457
2.8 -0.723999406346689
2.9 -0.739239372903425
3.0 -0.753962629724132
3.1 -0.768203068838742 /

\setdashes
\setquadratic 
\plot
2.3 -1.42192872636700
2.4 -1.44041213206101
2.5 -1.45814089902144
2.6 -1.47517423832022
2.7 -1.49156465450839
2.8 -1.50735892169162
2.9 -1.52259888824836
3.0 -1.53732214506907
3.1 -1.55156258418368 /

\endpicture
\caption{The $SU(3)\/$ Creutz ratios $\chi(i,j)$,
the scaling predictions for the string tension $\sigma a^2$, and
the tree-level curves
for $\chi(2,2)$ (dots) and $\chi(4,4)$ (dashes)
are plotted against $\beta$.
For $1.9 < \beta < 2.1$,
some of the symbols of the $\chi$'s overlap. }
\end{figure}
\smallskip
\section*{ACKNOWLEDGMENTS}
\par
We are indebted to
M.~Creutz,
G.~Marsaglia, F.~Palumbo, W.~Press, and K.~Webb
for useful conversations,
to the Department of Energy for financial support
under task B of grant DE-FG03-92ER40732/A004, and to
B.~Dieterle and the Maui Center for High-Performance
Computing\footnote[1]
{Research sponsored in part by the Phillips Laboratory, Air
     Force Materiel Command, USAF, under cooperative agreement
     F29601-93-2-0001.  The U.S. Government retains a
     nonexclusive copyright to this work.  The views and conclusions
     of this work are those of the authors and
     should not be interpreted as necessarily representing the
     official policies or endorsements, either expressed or
     implied, of Phillips Laboratory or the U.S. Government.}
for computer time.
\bibliographystyle{unsrt}

\end{document}